\documentclass[aps,amsmath,amssymb,preprintnumbers,nofootinbib,a4paper,prl,twocolumn]{revtex4-1}
\pdfoutput=1
\usepackage{placeins}
\usepackage{xspace}
\usepackage{amssymb,url}
\usepackage{graphicx}
\usepackage{hyperref}
\usepackage{color}

\usepackage{array}
\usepackage{amsmath}
\usepackage{mathtools}
\usepackage{epsfig}
\usepackage{multirow}
\usepackage{todonotes}
\usepackage[ margin=17pt, font=normalsize,labelfont=bf,justification=raggedright]{caption} 

\definecolor{rossoCP3}{cmyk}{0,.88,.77,.40}

\def\beq{\begin{equation}} 
\def\eeq{\end{equation}}
\DeclareMathOperator{\Tr}{Tr}

\begin{document}

\title{\Large  \color{rossoCP3} Preheating in an Asymptotically Safe Quantum Field Theory} 

\author{Ole Svendsen}
\email{svendsen@cp3-origins.net} 
\affiliation{
\vspace{3mm} 
{\rm CP}$^{\bf 3}${\rm-Origins},  
University of Southern Denmark, Campusvej 55, DK-5230 Odense M, Denmark}
\author{Hossein Bazrafshan Moghaddam}
\email{bazrafshan@physics.mcgill.ca}
\affiliation{
\vspace{3mm} 
Department of Physics, McGill University, Montreal, QC H3A 2T8 Canada}
\author{Robert Brandenberger}
\email{rhb@physics.mcgill.ca}
\affiliation{
\vspace{3mm} 
Department of Physics, McGill University, Montreal, QC H3A 2T8 Canada\\
and\\
Institute for Theoretical Studies, ETH Z\"urich, CH-8092 Z\"urich, Switzerland}

\begin{abstract}

We consider reheating in a class of asymptotically safe quantum
field theories recently studied in  \cite{Litim:2014uca, Litim:2015iea}.
These theories allow for an inflationary phase in the very early
universe. Inflation ends with a period of reheating. Since the models
contain many scalar fields which are intrinsically coupled to the inflaton
there is the possibility of parametric resonance instability in
the production of these fields, and the danger that the induced
curvature fluctuations will become too large. Here we show that
the parametric instability indeed arises, and that hence the energy
transfer from the inflaton condensate to fluctuating fields is rapid.
Demanding that the curvature fluctuations induced by the parametrically
amplified entropy modes do not exceed the upper observational bounds puts a
lower bound on the number of fields which the model of 
Ref.~\cite{Litim:2014uca, Litim:2015iea} must contain. This bound also
depends on the total number of e-foldings of the inflationary phase. 
  
{\footnotesize  \it Preprint: CP$^3$-Origins-2016-011 DNRF90 \& DIAS-2016-11}
 \end{abstract}

\maketitle
 
\section{Introduction}

Recently, interesting results have come up in the area of asymptotically safe quantum field 
theories \cite{Litim:2014uca, Litim:2015iea}, where it has been shown that a gauge-Yukawa 
theory can exhibit an ultraviolet (UV) safe (i.e. non-trivial) fixed point. Being UV safe these 
models could be interesting for early time cosmology. In particular, it has been
realized that these models admit a period of cosmological inflation at very early 
times \cite{Nielsen:2015una}.

In order to connect an early inflationary phase with late time cosmology, a period
of ``reheating" at the end of inflation is required. During the reheating phase, the
energy which is stored in coherent oscillations of the inflaton field at the end of
the period of inflation is transferred to a thermal bath of Standard Model particles,
yielding the onset of the post-inflationary radiation phase of cosmological expansion.
Reheating was initially studied using lowest order perturbation theory \cite{initial}.
This process is typically slow and leads to a low initial temperature of the
radiation phase. 

However, as was realized in \cite{TB, DK}, there may be a parametric resonance
instability which leads to a very rapid transfer of energy from the inflaton
condensate to fluctuations of the fields to which the inflaton couples. This
initial phase of energy transfer is called ``preheating" \cite{KLS1} and was
studied in detail in \cite{STB, KLS2}.  (see e.g. \cite{ABCM} and \cite{Karouby}
for recent reviews).

The preheating process typically leads to a non-thermal state in which modes in
certain wavelength intervals are highly excited whereas the rest are not
excited at all. However, it leads to
a phase in which the equation of state of matter is approximately that
of radiation. For early universe considerations such as baryogenesis
or the production of topological defects it is important to know the
energy density when the radiation phase of expansion begins. Hence,
it is important to know whether the preheating process is operative or not 
\footnote{Preheating does not arise in all inflationary models.}. The first
motivation for our study is to find out whether in the asymptotically safe
quantum field models discussed in \cite{Nielsen:2015una} preheating
occurs.

If an inflationary universe model admits a preheating instability at
the end of the period of inflation, there is the danger that the instability
will also affect the cosmological perturbations \cite{BKS1}. The period
of inflation produces fluctuations \cite{Mukh} which have the right spectral
shape to explain the observed distribution of matter in the universe and
the observed cosmic microwave background anisotropies. A 
parametric growth of these fluctuations at the end of inflation would
destroy the agreement between theory and observation. It was
shown that in a model with only a single scalar matter field, there
is no parametric amplification of the curvature fluctuations during
reheating \cite{FB1, AB}. However, in the presence of a second
scalar field a preheating instability of curvature fluctuations is
possible \cite{BV, FB2}. If we denote the inflaton field by $\varphi$
and the second scalar field by $\chi$, then it was found that
if the $\chi$ field experiences a parametric resonance instability
due to the coupling to the inflaton field (which is oscillating at the
end of inflation), that then an entropy fluctuation is generated
which leads to a growing curvature perturbation, even on length
scales which are much larger than the Hubble radius.

 It is important to point out why the growth of super-Hubble scale
  perturbations during reheating is compatible with causality. This is
  already discussed in [14].The key point is that in inflationary cosmology there is an exponentially large difference between the horizon 
(the forward light cone of a point on the initial condition surface, e.g. a point at the beginning of inflation) 
which grows exponentially in time, and the Hubble radius, the inverse expansion rate (which is constant during a period of exponential inflation). 
The reason why inflation can provide a solution of the horizon problem of Standard Big Bang cosmology is precisely the fact that the physical scale of a region of causal contact 
and homogeneity expands exponentially and becomes much larger than the Hubble radius. At the end of inflation, the inflaton field is in the coherent state and 
the correlation length of the field would set the maximum wavelength of fluctuations which can causally be amplified. 
Inflation will provide that the background inflaton field is coherent over a distance much larger than the Hubble radius during reheating and therefore guarantees
possibility of causal amplification of super Hubble modes. This possibility is the case for both adiabatic and entropy fluctuations. 
Also we would like to mention the fact that field equations are relativistic, hence causality is mathematically build-in and the result from field equations
will not violate causality. 
In this regard there is no distinction between supper-Hubble and sub-Hubble modes. For further details one may refer to section V of \cite{KBTM}.

The asymptotically safe quantum field models studied in \cite{Nielsen:2015una}
contains many scalar fields. A second goal of our study is
to see whether there is a parametric amplification of entropy
modes in our models, and what the resulting amplitude of the
induced curvature fluctuations is \footnote{This question has
recently been addressed in other two field models of inflation in
\cite{Keshav, Moghaddam:2014ksa, Moghaddam:2015ava, Axion}.}.

We find that there is indeed a parametric amplification of the
fluctuations of the $\chi$ fields in our model, and that this leads
to induced curvature fluctuations which grow on super-Hubble
scales during reheating. However, in the case in which the
model has a very large number of scalar fields (this is the
limit in which calculations leading to the presence
of the ultraviolet safe fixed point are under good controle) we
find that back-reaction effects shut off the instability before
the induced curvature fluctuations become too large.
 
\section{Model}

We will here give a short recap of the model at hand.
The full Lagrangian density is composed of  adjoint $SU(N_c)$ gauge fields, 
$N_f$ Dirac fermions in the fundamental of $SU(N_c)$ and an $N_f\times N_f$ 
neutral  complex scalar matrix, $H$.
We will here only present the scalar part of the Lagrangian.
 \begin{equation}
{\cal L} \, \supset \, 
 \Tr \left(\partial_\mu H^\dagger\partial^\mu H\right)-u\Tr\left(H^\dagger HH^\dagger H\right)-v\left(\Tr H^\dagger H\right)^2 \, ,
 \label{orglag}
 \end{equation}
where $u$ and $v$ are dimensionless coupling constants.

As we will be working in the UV regime of this model it is noteworthy that the 
scalar couplings at the UV fixed point \cite{Litim:2014uca} are given by 
\begin{eqnarray}
\alpha_u^* &=& \frac{u^*N_f}{(4\pi)^2} = \frac{\sqrt{23}-1}{19}\delta\label{fpu}\\
\alpha_v^* &=& \frac{v^* N_f^2}{(4\pi)^2} =-\frac{1}{19}\left( 2\sqrt{23}-\sqrt{20+6\sqrt{23}}\right)\delta
\label{fpv}
\end{eqnarray}
where the constant
\begin{equation}
\delta \equiv \frac{N_f}{N_c}-\frac{11}{2}
\end{equation}
(which must be positive) can be made arbitrarily small by adjusting $N_c$ and $N_f$.

We will in this work take a simplified version of \eqref{orglag} as we assume $H$ to be symmetric and real.
The parametrization of $H$ is given by
\beq
H_{ij} = 
\begin{cases}
   \frac{1}{\sqrt{2N_f}}\phi & \text{if } i=j \\
   \frac{1}{2}\chi_{(ij)}  & \text{else} \, ,
  \end{cases}
  \label{defh}
\eeq
where $(ij)$ indicates that this part is symmetric in $i$ and $j$.

With this normalization the kinetic term in \eqref{orglag} is given by
\beq
\Tr\left(\partial_\mu H^\dagger\partial^\mu H\right)=\frac{1}{2}\partial_\mu \phi\partial^\mu\phi+\frac{1}{2}\sum_{l}^{N_p}\partial_\mu\chi_{l}\partial^\mu\chi_{l}
\eeq
where $N_p=N_f(N_f-1)/2$ is the number of different off-diagonal fields, $\chi_{ij}$. It is clear 
from the kinetic term why we chose the normalization of $H$ in \eqref{defh}.
Similarly the double trace potential is given by
\beq \label{dtrace}
v\left(\Tr H^\dagger H\right)^2= \frac{v}{4}\phi^4+\frac{v}{4}\sum_l^{N_p}\chi_l^4+\frac{v}{2}\phi^2 \sum_l^{N_p}\chi_l^2+\frac{v}{2}\sum_{l>k}^{N_p}\chi_l^2\chi_k^2.
\eeq

For a general complex matrix $H$ the potential can be fully written in terms of the structure constant of $U(N_f)$, see Appendix B of Ref.~\cite{Rischke:2015mea}.
We will however here use the following grouping of terms for the single trace potential
\begin{equation} \label{strace}
\begin{split}
&u\Tr H^\dagger HH^\dagger H =\frac{u}{4N_f}\phi^4+\frac{3u}{2N_f}\phi^2 \sum_l^{N_p}\chi_l^2\\
&+\frac{3u}{\sqrt{2N_f}}\phi\sum_{i< j<k}^{N_f}\chi_{ij}\chi_{jk}\chi_{ki}+\frac{u}{8}\sum_l^{N_p}\chi_l^4\\
&+\frac{u}{8}\sum_{i\neq j\,\, k>i}^{N_f}\chi_{ij}^2\chi^2_{jk}+\frac{u}{16}\sum_{i,j,k,l}^{N_f}\chi_{ij}\chi_{jk}\chi_{kl}\chi_{li}.
\end{split}
\end{equation}
It is worth noting that there is no cubic term for $\phi$. This is a consequence of the fact
that $\phi$ appears only on the diagonal of $H$.

At first glance this model seems overly complicated, however a short motivation why this model is relevant for discussion will be given here. 
A central parameter in the study for parametric resonance is the ratio of the quartic inflaton coupling ($\lambda$) to the portal coupling ($g^2$).  
A study similar to the present can be given in a toy model with an inflaton and a scalar field coupled to the inflaton. In these toy models the relevant 
parameter ($\lambda/g^2$) will need to be fixed arbitrarily, see e.g. \cite{KLS1, KLS2, Moghaddam:2014ksa, Moghaddam:2015ava}. 
This is in contrast to the model of this paper. With the model given by \eqref{orglag} this ratio is given by the model it self and is thereby not an arbitrarily chosen number. 
Furthermore, as was shown in \cite{Litim:2015iea} the running of the couplings follow that of the gauge coupling (in \cite{Litim:2015iea} called  $\alpha_g$) along 
the UV uni-dimensional stable trajectory. This implies that the ratio $\lambda/g^2$ stays constant even including running away from the UV fixed point. 
This is a remarkable fact that solidifies our future choice for this ratio to be that at the UV fixed point. This is a priori not a feature of a toy model,
hence the model in  \eqref{orglag} is an interesting scenario.

\section{Recap of Results on Inflation}

We will in this section recap some of the results of Ref.~\cite{Nielsen:2015una} as this is 
the inflationary scenario we have in mind for our investigation. 
Inflation is driven by the diagonal element of $H$ where all diagonal elements are taken 
to be the same \cite{Litim:2015iea}. The inflationary effective potential can be derived 
from \eqref{orglag} with the couplings given by \eqref{fpu} and \eqref{fpv},
\beq
V(\phi) = \frac{\lambda\phi^4}{4 (1+W(\phi))}\left(\frac{W(\phi)}{W(\mu_0)}\right)^\frac{18}{13\delta}
\label{runningpot}
\eeq
where $\lambda=v^* +\frac{u^*}{ N_f}$ and $W(\cdot)$ is related to the product logarithm, see Ref~\cite{Nielsen:2015una} for details. This is a $\phi^4$-theory including renormalisation of 
the inflaton operator expanded near the UV fixed point.

The potential in equation \eqref{runningpot} is valid for a large range in $\phi$. However, for the study of parametric resonance small 
field values are considered as this minimum of the potential is located here.  This means that for this analysis the potential will 
be approximated by $\lambda\phi^4$. This full $\phi$ dependence is included for completeness. 

It was shown that this model can provide a viable scenario for large field inflation. The
inflationary slow-roll condition ceases to be satisfied and thus 
quasi-exponential expansion stops at a field value
\beq \label{phiendmodel}
\phi_{end} =\sqrt{(4-\frac{16}{19}\delta)(3-\frac{16}{19}\delta)}M_p\simeq\sqrt{12}M_p.
\eeq
The phenomenological predictions for this model, for very small $\delta$, lie just outside 
the Planck'15 $2\sigma$ contours for the tensor-to-scalar ratio and scalar spectral index. 
It was noted that in this perturbative regime a very large number of flavors was needed to 
produce the measured amplitude of curvature perturbations by the inflaton fluctuations alone. 
However this number drops rapidly as the perturbative parameter $\delta$ is pushed 
close to and beyond the radius of convergence of the underlying model. 

The inflationary phase will quasi-exponentially redshift the wavelength of
any fluctuations existing before the onset of inflation, and will produce a homogeneous
inflaton condensate. Once the field value of this condensate decreases to below
the value given by \eqref{phiendmodel}, accelerated expansion of space ends and the
reheating period begins. We will be working in terms of the usual metric
\beq
ds^2 \, = \, dt^2 - a(t)^2 d{\bf x}^2 
\eeq
of space-time, where $t$ is physical time and ${\bf x}$ are the Euclidean
comoving coordinates of the expanding space. It will often be useful to work
in terms of conformal time $\eta$ defined via $dt = a d\eta$. Since we
are interested in the period of reheating, we will normalize the scale
factor to be $a(t_R) = 1$ at the time $t_R$ corresponding to the beginning
of reheating.

\section{Parametric Resonance}

We will here discuss parametric resonance of the inflaton and the off-diagonal scalar fields. 
One could also investigate parametric production of fermionic fields; however this will be left 
for a later discussion. 

At the end of the inflationary epoch and before significant production of any other field has 
happend, the equation of motion (EoM) for the homogeneous inflaton field is given by
\beq
\ddot{\phi} + 3H\dot{\phi} + \lambda\phi^3 = 0 \, ,
\eeq
where $H$ is the Hubble expansion parameter. It is obvious that the solution
for $\phi$ will correspond to damped oscillation. The expansion of
space can be factored out by introducing a rescaled field
$\tilde{\varphi} \equiv a \phi$ and working in terms of conformal time. 
In terms of this field the solution is,
as noted in \cite{Greene:1997fu}, oscillatory, however not sinusoidal but 
proportional to the elliptic cosine, $cn(x)$, where $x$ is a rescaled dimensionless
conformal time which will be defined below. The equation of state of
these oscillations is (upon time averaging) that of radiation. Hence the
amplitude, which is asymptotically given by
\beq
\tilde{\varphi} = a \frac{1}{\sqrt{t}}\left(\frac{3M_P^2}{8\pi \lambda}\right)^\frac{1}{4} \, ,
\eeq
is constant.

The EoM for the $ij$ off diagonal component is given by
\begin{eqnarray}
0=&&\ddot\chi_{ij}+3H\dot\chi_{ij}+g^2\phi^2\chi_{ij} - a^{-2} \nabla^2 \chi_{ij} \nonumber \\ 
&&+\frac{3u}{\sqrt{2N_f}}\phi\sum_k^{N_f}\chi_{ik}\chi_{kj}+\mathcal{O}(\chi^3) \, ,
\end{eqnarray}
where $\nabla$ stands for the gradient operator with respect to the
comoving spatial coordinates. Here 
\beq
g^2 = v +\frac{3u}{N_f} \, .
\eeq
Note that first line is leading order in $\chi\sim0$, and anything beyond this is subleading.
In our study of parametric resonance we consider the first three terms only, and the 
rest we regard as back reactions in the next section. The first line is a linear
equation for $\chi_{ij}$ which we will solve for earch Fourier mode independently.

Fourier transforming this linear EoM and rescaling the field by $a\chi =X$ yields \footnote{We are neglecting terms in the equation of motion for $X_k$
containing the metric perturbations. As discussed in
Chapter 19 of the review article \cite{MFB} this is justified
as long as the energy density in the entropy field is smaller
than that of the inflaton field. We are not the first to use
this approximation. It is the basis of the {\it curvaton
scenario} \cite{curvaton} and related scenarios such as
the \it{New Ekpyrotic model}  \cite{NewEkp}.}
\beq
\tilde X_k'' + \left(\kappa^2 
+ \frac{g^2}{\lambda}cn^2\left(x,\frac{1}{\sqrt{2}}\right)-\frac{a''}{a}\right)\tilde X_k = 0 \, ,
\label{dimlesschieom}
\eeq
where we have dropped the $ij$ indices since the equation is identical for
each component as long as we do not include mixing terms from the higher orders in $\chi$. 
We have defined
\beq
\kappa^2 = \frac{k^2}{\lambda\tilde{\varphi}^2}
\eeq
and primes are derivatives w.r.t the dimensionless conformal time
\beq
x \equiv \sqrt{\lambda}\tilde\varphi \eta \, . 
\eeq
In \eqref{dimlesschieom} we have signified the Fourier component by a tilde, however they will be omitted from now on.
The term $\frac{a''}{a}$ is zero during the phase of parametric resonance as can be checked from 
the Friedman equation. 

We will now approximate the elliptic cosine by the first term of a cosine expansion 
\cite{Greene:1997fu} as follows
\beq
 cn\left(x,\frac{1}{\sqrt{2}}\right)\simeq \frac{8\sqrt{2}\pi}{T}  \frac{e^{-\frac{\pi}{2}}}{1+e^{-\frac{\pi}{2}}}\cos\left(\frac{2\pi}{T}x\right)
 \eeq
 where T is giving by the complete elliptic integral 
 \beq
 T = 4K\left(\frac{1}{\sqrt{2}}\right)=4F(\frac{\pi}{2}\vert \frac{1}{2} ) \, . \nonumber
 \eeq
 Here 
 \beq
 F(\theta\vert m^2) = \int^\theta_0\frac{d\phi}{\sqrt{1-m^2 \sin^2(\phi)}} \, \nonumber
 \eeq
 is the elliptic integral. 
 Numerically we have $\beta \equiv \frac{2\pi}{T} \simeq 0.8472$.
 
This series expansion is not ideal for our situation. It might be the best behaved in terms of convergence, however for approximate use the amplitude of the first term is too small. 
As an approximation to the elliptic cosine we will use
 \beq
 cn\left(x,\frac{1}{\sqrt{2}}\right)\simeq  \cos\left(\beta x\right).
 \eeq 
With this we can rewrite \eqref{dimlesschieom} in terms of the rescaled dimensionless 
conformal time, $y \equiv \beta x$ (which we normalize to be $y = 0$
at the beginning of the reheating period, i.e. $t = t_R$) as
\beq
X_k'' + \beta^{-2}\left(\kappa^2 + \frac{g^2}{2\lambda} + 
2\frac{g^2}{4\lambda}\cos\left(2y\right)\right)X_k = 0
\eeq
where the derivative is with respect to $y$.

Written on this form, the equation of motion for $\chi$ takes the form of a Matheiu equation with parameters $q=\frac{g^2}{4\beta^2\lambda}$ and $A=\beta^{-2}(\kappa^2+2q)$
(see e.g. \cite{Mathieu}). 
The Matheiu equation has a solution given by
\beq
X(y) = A_1\exp(\mu_k y) P_1 (y,k)+A_2\exp(-\mu_k y) P_2 (y,k),
\eeq
that is, a growing and decaying exponential solution, with periodic behavior captured 
by $P_1$ and $P_2$. Here $\mu_k$ is the Mathieu characteristic exponent, and the 
two periodic functions have amplitude $1$ 
and the frequency  which is given by the frequency of the inflaton condensate, i.e. independent of 
$k$ (see \cite{Mathieu, Floquet}).

\begin{figure}
\centering
\includegraphics[width=\columnwidth]{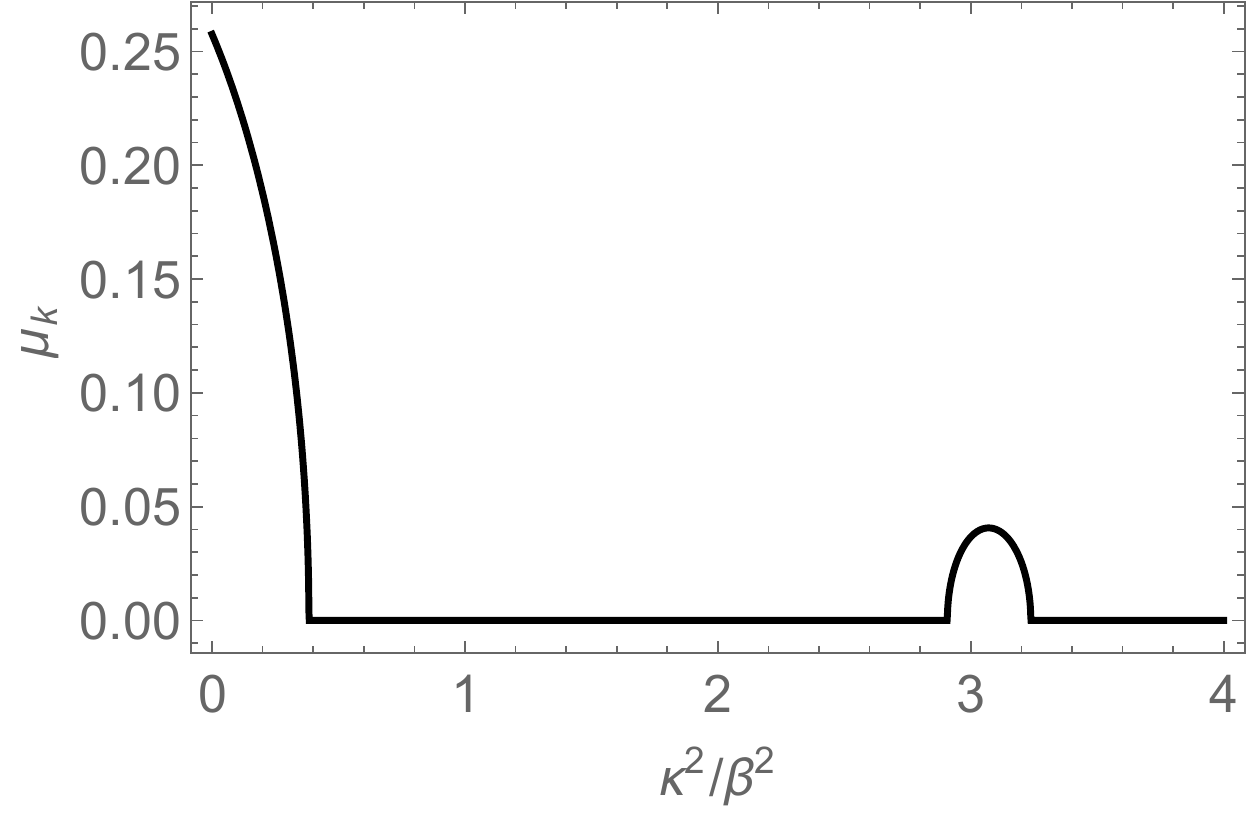}
\caption{Positive real part of the Mathieu characteristic exponent for 
$q=\frac{g^2}{4\beta^2\lambda}$ and $A=\beta^{-2}(\kappa^2+2q)$.}
\label{matexp}
\end{figure}

The characteristic exponent $\mu_k$ is in general a complex number. However, for some parameters it 
has a real part called the Floquet index. Given our model parameters
\beq
 \frac{g^2}{\lambda} = \frac{\alpha_v^*+3\alpha_u^*}{\alpha_v^*+\alpha_u^*} \simeq 7.4 \, ,
\eeq
we present the Floquet index in Figure \ref{matexp}.

Before discussing the parametric amplification of $\chi$ after
inflation, we must determine the initial conditions for $\chi(y)$ at the end of
the inflationary phase. The fluctuation modes of the rescaled variables $X$ (which
are the canonical variables) begin in their quantum vacuum state. We now need
to investigate whether or not fluctuation in $\chi$ are squeezed during inflation on
super-Hubble scales (on sub-Hubble scales the $k^2$ term  prevents
squeezing in the same way that the corresponding term prevents the
squeezing of inflaton fluctuations). Squeezing of the $X$ fluctuations
will transform the initial vacuum spectrum into a scale-invariant one
in the same way that it leads to a scale-invariant spectrum of super-Hubble inflaton
fluctuations.  To check whether there is squeezing of the $X$ perturbations 
during inflation we compare the neglected $\frac{a''}{a}$ term from \eqref{dimlesschieom}
to the induced mass term $\frac{g^2}{\lambda}$.
In terms of  regular time the condition for squeezing becomes
\beq
2H^2 > g^2 \frac{\tilde\varphi^2}{a^2} = g^2 \phi^2 \, .
\eeq
Expressing the value of $H$ during slow-roll inflation in terms of the
field $\phi$ show us that the squeezing condition is satisfied throughout
the period of inflation. Hence the spectrum of the $X$ perturbations 
will be scale invariant at the beginning of the reheating period ($t = t_R$), i.e.
\beq
X_k(t_R)  \simeq H_I k^{-3/2} P_1(y = 0,k) \, ,
\label{xevol}
\eeq
where $H_I$ is the value of $H$ during inflation (more precisely when
the scales of interest exit the Hubble radius during inflation). In
the following we will not make a difference between $H_I$ and the
Hubble expansion rate at the end of the inflationary phase, i.e. $H(t_R)$.

During the preheating phase, the above value of $X$ is exponentially
amplified, yielding
\beq 
X_k \simeq H_I k^{-3/2} \exp(\mu_k y) P_1(y,k) \, .
\label{xevol}
\eeq
Since we have normalized the scale factor to be $a(t_R) = 1$, and since
we can ignore the growth of $a(t)$ during the initial preheating period,
we can identify $X$ and $\chi$.

The fluctuations of $\chi_k$ computed above yield the entropy fluctuations
generated in our model. In order to compute the induced curvature
fluctuations, we need the background value of the $\chi$ field. 
In the
case of two field inflationary models with scalar fields having classical
background, it is clear how to identify the background value of the
entropy field. In the case of a model like the one we are considering,
in which there is no classical zero mode of the entropy field, the
situation is more complicated. Working strictly at first order in
perturbation theory there is no background, and hence there will be
no induced curvature fluctuations. However, this is clearly not the
correct result, since if we were to argue in this way then cosmic
string formation in an early universe phase transition would never
lead to curvature fluctuations on super-Hubble scales, and it is well
known that such curvature perturbations are formed (see e.g. \cite{CSrevs}
for reviews on cosmic strings and structure formation). A way
to address this issue was recently suggested in 
\cite{Moghaddam:2014ksa, Moghaddam:2015ava, Axion}: 
each $k$ mode of the fluctuations lives in an 
{\it effective background} $\chi_{ij}^{eff}(k)$ which is generated by all perturbation 
modes with smaller wavelengths. 

The effective background is given by \footnote{To further justify this, imagine that $\chi$ contains fluctuations
with two Fourier modes, a mode $k$ we are interested
in, and a longer wavelength
mode $k'$. The mode $k$ can locally be viewed as a mode
which fluctuates not about $0$, but about the local value
of the $k'$ mode.}
\beq \label{chieffdef}
\chi_{ij}^{eff}(k) = \left(\int_0^k d^3k'|X_{k'}|^2\right)^{1/2} \, .
\eeq
To see this, we begin from the expression for the contribution of
long (i.e. longer than $k^{-1})$ modes to the $X$ field
at a fixed point $x$ in space 
\beq \label{Xlong}
X(x) \, = \, V^{1/2} \int_0^k d^3k' X_{k'} e^{i k' x} \, ,
\eeq
in terms of the Fourier modes. Here, $V$ is the cutoff spatial volume
which we introduced such that the Fourier modes have the mass
dimension of a harmonic oscillator. Without loss of generality we
can take the point $x$ to be $x = 0$. We now consider the
expectation value of the square of the absolute value of (\ref{Xlong})
\beq \label{square}
< |X(0)|^2 > \, = \, V \int_0^k d^3k' \int_0^k d^3k'' <X_{k'} X^{*}_{k''} > \, .
\eeq
Since the Fourier modes are uncorrelated we have
\beq \label{expvalue}
<X_{k'} X^{*}_{k''} > \, \, \delta^3(k' - k'') V^{-1} |X_{k'}|^2 \, .
\eeq
Inserting (\ref{expvalue}) into (\ref{square}) then yields (\ref{chieffdef}).

We are interested in infrared modes $k$ which lie in the instability band
of the Mathieu equation. From Figure \ref{matexp} we see that for these
values of $k$ the infrared modes $k'$ which appear inside the integral 
also lie in the instability band, and that we can approximate the Floquet
exponent by a constant $\mu_{k'} = \mu$. Inserting (\ref{xevol}) into
(\ref{chieffdef}) we see that there is a potential logarithmic infrared divergence of
the integral. The periodic function $P(y, k')$ appears in the
integrand in quadratic form and hence does not eliminate the
infrared divergence. There is, however, an infrared cutoff: the form (\ref{xevol})
does not apply for modes which are outside the Hubble radius at the
beginning of inflation.  Therefore the integral can be estimated by
\beq
\chi_{ij}^{eff}(k) \sim  H_I \exp(\mu y) \left( {\rm ln}\left(\frac{k}{k_{min}}\right) \right)^{\!1/2} 
P_1(y, k) \, ,
\label{chieff}
\eeq
where $k_{min}$ is the value of $k$ which corresponds to Hubble
radius crossing at the beginning of the period of inflation. The logarithm
is given by $N_I$, the number of e-foldings of inflation.

\section{Back Reaction Effects}

In the previous section we studied preheating in the UV-safe theory we
introduced in the current paper neglecting any back-reaction mechanism.
Having done that we observed that exponential amplification sets in,
and that the induced curvature fluctuations might have potential dangerous 
consequences for the theory. However it is crucial to consider back-reaction 
effects since they will eventually terminate the resonance. Back-reaction
effects are nonlinear and are often studied numerically. However, for
the questions of large-scale curvature fluctuations, an analytical
analysis is preferable (see \cite{Zibin} for an initial study of
back-reaction effects during preheating in a two field inflation model).

Here we will consider two kinds of back-reaction effects: 
\begin{itemize}
\item The effect of produced $\chi$ particles on the evolution of the inflaton
field .
\item The contribution of amplified modes on the effective mass for $\chi$
field fluctuations.
\end{itemize}
Other back-reaction effects are studied in \cite{Greene:1997fu,Moghaddam:2015ava} 
but are not relevant in our case.

\subsubsection{Effects on the evolution of the inflation field}

The leading order term in $\chi$, for small $\chi$, in the inflaton action is 
\beq
\frac{g^2}{2} \phi^2 \sum_l^{N_p} \chi_l^2 \, ,
\eeq
where the index $l$ runs over all of the $\chi$ fields.
This must remain sub-dominant to the main interaction term in the inflaton
action which is
\beq
\frac{\lambda}{4}\phi^4 \, . 
\eeq
This leads to the condition
 \beq
\sum_l^{N_p} \langle\chi_l\rangle_{eff}^2 < \frac{\lambda}{2g^2}\phi_{end}^2 \, 
 \eeq
which needs to be satisfied in order to justify neglecting
back-reaction effects. In the case when all preheat fields are excited equally, 
which we show is the case, we get
\beq
\langle\chi_l\rangle_{eff}^2 < \frac{\lambda}{2g^2 N_p}\phi_{end}^2  \, .
\label{relevantbr}
 \eeq
Naturally, the more fields we have, the less each must be excited to interfere. Note
that all $\chi$ modes which are excited contribute to the left hand side of
(\ref{relevantbr}), and hence the expression is proportional to $N_I$.

\subsubsection{Contribution to the effective mass of the preheat field fluctuation}
 
Now we focus on the mass term in the equation of motion for one of the
$\chi_k$ field modes. At linear order in $\chi$, the mass term comes from
the coupling of $\chi$ to the inflaton. This is the mass term which we
have considered and which leads to the parametric instability, and
its value is
\beq
 g^2 \phi_{end}^2 \, .
 \eeq
However, beyond linear order there is a contribution to the mass which comes from 
the interactions between all $\chi$ fields. As is evident from
(\ref{dtrace}) and (\ref{strace}) each $\chi$ field couples quadratically
to a fixed $\chi$. Assuming that all $\chi$ fields are excited 
equally we get a contribution to the mass which is
\beq
\lambda' N_p \langle\chi\rangle_{eff}^2
 \eeq
where we have taken into account that each mode of $\chi$ which
are excited contribute to the effective mass of the $\chi$ field,
and hence the above expression is proportional to $N_I$. In the above,
$\lambda'$ is a coupling constant made up of the constants appearing in
(\ref{dtrace}) and (\ref{strace}). The condition, that back-reaction can
be neglected, then becomes
\beq \label{brcond2}
\langle\chi\rangle_{eff}^2 < \frac{ g^2}{2\lambda' N_p}\phi_{end}^2 .
\eeq
Since $\lambda'$ is of the same order of magnitude as $\lambda$
we find that (using the value of $g^2 / \lambda$ which our model
predicts) the first back-reaction condition (\ref{relevantbr}) is slightly 
stronger than the second one (\ref{brcond2}).
 
Note that when the parametric resonance stops all preheating fields have 
been excited equally as the sole non-democratic couplings enter in the $\chi^4$ 
terms relevant only for the second back-reaction effect considered.

\section{Induced Curvature Perturbations}
 
Having shown that parametric resonance of the
spectator scalar fields is efficient in this model, we 
move on to investigate the resulting amplification of 
the entropy fluctuations which in turn leads to a contribution 
to the curvature perturbation with an exponentially growing 
amplitude.   

Fluctuations in a spectator scalar field
will induce a contribution to the curvature whenever the
equation of state of the spectator field mode is different
from that of the adiabatic mode. The magnitude
of the induced curvature fluctuation is proportional
to the energy density in the spectator field. Thus, a
background value of the spectator field is required
in order to obtain a growing curvature mode. If we
are considering curvature fluctuations on a fixed 
scale $k$, we will use the effective background $\chi$
field constructed earlier as this background.  
The conversion of entropy field fluctuations into curvature
perturbations has been studied in many works (see e.g.
\cite{Gordon, Malik} for some classic papers). We may 
also use the covariant formalism of \cite{Ellis:1989jt} as applied 
to study preheating in \cite{Moghaddam:2014ksa}. 
The result of these analyses is that the induced
curvature fluctuation is given by
\beq
\zeta _k \simeq \frac{H}{\dot\phi^2}\dot\chi S_k	 \, ,
\label{oldresult}
\eeq
where $S_k$ is the entropy field perturbation
which, according to our analysis, is given by
\beq
S_k = H k^{-3/2} \exp (\mu y) P_1(y,k) \, ,
\eeq
where $P_1(y,k)$ is the same periodic function as in \eqref{xevol}.

The expression for $\dot{\chi}$ can be found using \eqref{chieff}, yielding
\beq
\dot \chi = \beta \sqrt{\lambda}\Phi H \sqrt{N_I} \exp (\mu y) \frac{\partial P_1(y,k)}{\partial y} \, ,
\eeq
where $\Phi$ is the amplitude of $\phi$, and where
we have neglected the derivative of the exponential factor, as it carries 
a factor of $\mu$ which is small compared to the order $1$ frequency of $P_1$. 
Hence, we obtain
\beq
\zeta_k = \frac{H^3}{\dot\phi^2}\sqrt{\lambda} \beta \sqrt{N_I} \exp(2\mu y) k^{-3/2}  \Phi P(y,k) 
\eeq
where $P(y,k)=P_1 \frac{\partial P_1}{\partial y}$ is a periodic function. The most 
important feature of this result is exponential growth of curvature perturbation 
which is induced by the entropy perturbation (see also the Appendix of 
Ref~\cite{Moghaddam:2014ksa} for a more detailed derivation of this result).

Using this result we can evaluate the power spectrum of induced curvature 
perturbations, yielding 
\beq
P_k = \frac{k^3}{2\pi^2} \vert \zeta_k\vert^2 \simeq \frac{H^6 \Phi^2}{{\dot\phi}^4} \exp(4\mu y) 
\frac{\beta^2 \lambda}{4 \pi^2} N_I \, ,
\label{curper}
\eeq
where we used $\vert P \vert ^2 \sim \frac{1}{2}$. To estimate this expression we will 
use the fact that at the end of inflation kinetic energy is of the same order of magnitude 
as the potential energy. We also use $\Phi=\phi_{end}$ and introduce the number
$\sigma$ via
\beq
\phi_{end} \equiv \sigma M_p.
\label{phiend}
\eeq
We can use the result of the previous section on back-reaction to yield an estimate 
for the value of $\exp(4\mu y)$ when the resonance stops. As discussed in the 
previous section one can show the first back-reaction effect shuts off the
resonance before the second. Therefore using \eqref{relevantbr} for the time 
when the back-reaction becomes important, we get
\beq \label{endvalue}
\exp(4\mu y) \sim \left(\frac{2 g^2}{\lambda}\right)^{-2} \frac{M_p^4}{H^4} N_I^{-2} N_p^{-2} \sigma^4
\eeq
Inserting (\ref{endvalue}) and (\ref{phiend}) into (\ref{curper}) and taking into
account that the potential energy at the end of inflation is
\beq
V \, = \frac{\lambda}{4} \sigma^4 M_p^4
\eeq
we obtain our final result
\beq
P_k \sim \frac{\beta^2 \sigma^2}{N_p^2 N_I}\left(\frac{g^2}{\lambda}\right)^{-2}
\label{pwspec}
\eeq
for the power spectrum of the induced curvature fluctuations.

For our model with $\frac{g^2}{\lambda}\simeq7.4$ and $\sigma^2 \simeq 12$ 
and $\beta \simeq 1$ we get
\beq
 P_k \sim \frac{1}{N_p^2 N_I}.
\eeq
For this not to exceed the observed value with amplitude of order $10^{-10}$ we need
a large number of flavors and/or a large number of e-foldings of inflation.

\section{Conclusion}
 
In this paper we have considered reheating in a class of asymptotically safe quantum
field theories recently studied in  \cite{Litim:2014uca, Litim:2015iea}.
These theories allow for an inflationary phase in the very early
universe. Inflation ends with a period of reheating. Since the models
contain many scalar fields which are intrinsically coupled to the inflaton
there is the possibility of parametric resonance instability in
the production of these fields, and the danger that the induced
curvature fluctuations will become too large. 

Our first results is that the parametric instability indeed arises, and that hence the 
energy transfer from the inflaton condensate to fluctuating fields is rapid.

Our second result concerns the demand that the curvature fluctuations 
induced by the parametrically amplified entropy modes do not exceed 
the upper observational bounds. We have seen that this puts a lower bound on the product
$N_p^2 N_I$, where $N_p$ is the number of scalar fields 
which the model of  \cite{Litim:2014uca, Litim:2015iea} contains,
and $N_I$ is the total number of e-foldings of the 
inflationary phase. The reason that the power spectrum of the
induced curvature fluctuations decreases
as $N_p^2$ is that back-reaction effects turn off the parametric
instability earlier as $N_p$ increases. It is a linear effect in $N_p$
on the fluctuation modes, and hence a quadratic effect in the
power spectrum. The reason that the bound depends on
$N_I$ is that the energy density in the effective entropy field background (which
determines the strength of the conversion of entropy to adiabatic mode)
is proportional to $\sqrt{N_I}$, and that as $N_I$ increases the
back-reaction is shut off earlier due to more modes being super-Hubble. The combination of these effects
gives the net scaling of the power spectrum as $N_I^{-1}$.

Although we chose to investigate the parametric production of
entropy fluctuations in a specific particle physics model (which
was of interest to us for other reasons), our analysis can be extented rather straightforwardly to other multi field models. 
Features found in this paper will be present in other models as well, making our analysis more generally applicable.
 
 \vskip .2cm
 
 \section*{Acknowledgments} 
 
 \noindent
One of the authors (RB) wishes to thank the Institute for Theoretical
Studies of the ETH Z\"urich for kind hospitality. He acknowledges
financial support from Dr. Max R\"ossler, the Walter Haefner Foundation
and the ETH Zurich Foundation, and from a Simons Foundation fellowship.
The research of RB is also supported in part by funds from NSERC and the
Canada Research Chair program. 
The work of OS is partially supported by the Danish National Research Foundation grant DNRF:90.
HBM is supported in part by a MSRT fellowship.

\end{document}